\documentclass[12pt,preprint]{aastex}

\lefthead{Escala}
\righthead{Evidence in Favor of Direct Massive Black Hole Formation.}

\def\msun{M_{\odot}}

\begin{document}

\title{Observational Support  for   Massive Black Hole Formation   Driven by  Runaway Stellar Collisions   in Galactic Nuclei.}

\author{Andr\'es Escala}
\affil{Departamento de Astronom\'{\i}a, Universidad de Chile, Casilla 36-D, Santiago, Chile.}
\affil{aescala@das.uchile.cl}

\begin{abstract}

%\keywords{galaxies: formation - star formation: general}

{\bf We explore here
  an scenario for massive black hole  formation driven by  stellar collisions in galactic nuclei%We 
  %and hypothesized
  , proposing  a new formation regime  %for massive black hole  formation through 
  of   global instability in nuclear stellar clusters  triggered  by runaway stellar collisions%, in addition to the standard regime  driven by stellar collisions restricted to the inner core, that  leads  to the formation of  black holes with intermediate  masses
  . Using order of magnitude estimations, we show that observed nuclear stellar clusters avoid the regime where stellar collisions are dynamically relevant over the whole system, while resolved detections of massive black holes are well into such collision-dominated regime. We interpret this result in terms of massive black holes and nuclear stellar clusters being different evolutionary paths of a common formation mechanism, unified under the standard terminology of being both central massive objects. We propose a formation scenario where central massive objects more massive than $\rm \sim 10^8 \, \msun$, which also have relaxation times longer that their collision times,   will be too dense (in virial equilibrium) to be globally stable against stellar collisions and most of its mass will collapse towards the formation of a massive black hole. Contrarily, this will only be the case   at the core of less dense central massive objects leading  to the formation of  black holes with much  lower  black hole efficiencies  $\rm  \epsilon_{BH} = \frac{M_{BH}}{M_{CMO}}$, with  these  efficiencies   $\rm  \epsilon_{BH}$ drastically  growing for central massive objects  more massive than $\rm \sim  10^7 \,\msun$,  approaching    unity around $\rm M_{CMO} \sim 10^8 \, \msun$. We show that the proposed scenario successfully explains the relative trends observed in the masses, efficiencies, and scaling relations between massive black holes and nuclear stellar clusters.
}
\end{abstract}
\section{Introduction}

For more than half a century,  evidence was  accumulated for the existence of Massive Black Holes (MBHs)  in galactic nuclei with masses $\rm \sim 10^{6-9} \msun$ (Zel'dovich 1964; Salpeter 1964; Ghez et al. 2008; Gillessen et al. 2009), but only recently arrived definite support  in favor of their existence (The Event Horizon Telescope Collaboration 2019). The origin of such `monsters'  puzzled theorists soon after their discovery (Rees  1984), however,  is still a mystery  their dominant formation process (Volonteri 2010). With the advent of the gravitational wave astronomy, specially with the  future  LISA experiment (Amaro-Seoane  et al. 2013; Barause et al. 2015), it is expected to have definite answers on the formation of MBHs in the universe.

Several pathways have been proposed for MBH formation (Rees  1984; Volonteri 2010; Shapiro  2004), which can be briefly summarized into 3 channels: i) Direct collapse of a primordial cloud onto a MBH (Bromm \& Loeb 2003; Lodato  \& Natarajan  2006; Latif \& Schleicher 2015; Becerra et al. 2015; Regan \& Downes 2018) ii) Growth  by gas accretion and/or mergers  of a  stellar/intermediate mass BH up to the mass range of  MBHs (Madau \& Rees 2001; Volonteri et al. 2003; Agarwal et al. 2013; Ricarte \& Natarajan 2018) and iii) Formation of a %VMS/      
MBH by catastrophic stellar collisions in dense stellar clusters (Zel'dovich  \& Podurets 1965; Rees  1984; Shapiro  \& Teukolsky 1985; Omukai et al. 2008; Devecchi \& Volonteri 2009; Devecchi et al. 2014). However, all of them faced severe  problems to fulfill the constraints set by observations, such as the physical conditions needed to sustain   atomic cooling halos are unclear to be fulfilled (i; Shang et al. 2010; Inayoshi \& Haiman 2014; Suazo et a. 2019), problems for    explaining  the highest redshift  quasars by  lower mass BHs grown thru Eddington-limited  accretion (ii; Gnedin  2001; Volonteri \& Rees 2006; Prieto et al. 2017; Inayoshi et al. 2020) and that  simulations of stellar collisions in dense  clusters are able to form only BHs of lower masses in the intermediate mass regime (iii; Portegies Zwart \& McMillan 2002; Freitag et al. 2006a,b; Goswami et a. 2012; Stone et al. 2017). %(What scenarios don't work, one line with pro/cons (ver review Volonteri/LISA)). In ther absence of supporting mechanisms, you will ends up into a MBH.

%On the other hand, 
Besides the problems faced by the different  formation scenarios, galactic centers are arguably the most favorable places for MBH formation. Any gaseous  (and stellar) material  that  eventually losses its orbital support falls on to this preferential place (Shlosman et al. 1990; Escala 2006), which corresponds to the deepest part of the galactic  gravitational potential. Multiple  processes produces strong inflows  at galactic scales on dynamical timescales, funneling  large amount of gaseous material (up to  $\rm 10^{10}\, \msun$) to this preferential place%,  especially in the absence of efficient AGN feedback (Prieto 2017) Those processes 
, that includes   gravitational  torques in galaxy mergers (Barnes  2002; Mayer et al 2010; Prieto et al 2020), bars within bars (Regan \& Teuben 2004; Hopkins \& Quataert 2010), clump migration by dynamical friction (Escala 2007; Elmegreen et al. 2008), etc. These  processes  are expected to be even more dramatic in the case of proto-galactic material at high z,  because of  the higher gas fractions and the absence of AGN feedback  from preexisting MBHs (Prieto \& Escala  2016). Therefore, in the absence of %AGN 
feedback limiting factors (Dubois et al. 2015; Prieto et al. 2017),
  the amount of material funneled into galactic nuclei has (in principle) %a priori 
   no  upper limit  externally set by processes at galactic scales and thus, we expect to be the  hosting place of the densest gaseous and stellar configurations %(at least locally?)
 in the universe. The straightforward  question is then, if such material it does  not ends up forming  a MBH, that corresponds to gravity's final triumph, which other  stable physical configuration (at intermediate densities) it could be?

The hypothetical scenario under   very efficient heating mechanisms ($\rm T_{vir} \geq 10^4 K$) has been  extensively  studied (Bromm \& Loeb 2003; Lodato  \& Natarajan  2006; Latif \& Schleicher 2015; Becerra et al. 2015; Regan \& Downes 2018),  where  fragmentation is suppressed on smaller scales and that directly leads to the formation of a single Very Massive quasi-Star (VMS; Volonteri  \& Begelman  2010; Schleicher  et al. 2013) at the center, which  afterwards collapses   onto  a  MBH due to post-Newtonian instability (Tolman 1934; Oppenheimer \& Volkoff 1939). Contrarily, in the absence of efficient heating the gaseous material funneled to the galactic center  efficiently cools (Rees \& Ostriker  1977; Sarazin \& White 1987), eventually becomes unstable and fragments in a broad range of scales (Toomre 1964; Escala \& Larson  2008; Escala 2011), leading to the   formation of  a dense stellar cluster (Bate et al. 2003; Padoan et al. 2016). Such dense stellar configurations are indeed observed, being called   Nuclear Stellar Clusters (NSCs) and that are  considered  the densest stellar configurations  in the local Universe (Boker et al. 2004; Cote  et al.  2006; Walcher  et al. 2006; Balcells  et al. 2007), located in galactic nuclei and where   in some cases  coexists with a MBH  (Leigh et al. 2012; Georgiev et al 2016), thus having possibly a joint formation event.  Therefore, a more realistic   scenario %key question is now 
reduces  to how dense such stellar system it can be before becoming globally  unstable,  leading it again to the formation of a MBH.

 A natural   candidate   for triggering instability  in stellar clusters are collisions between stars since it is  an efficient  mechanism  for loosing  orbital energy   support, because physical collisions between stars  are a dissipative source on a fluid interpretation of a  cluster, being able to convert  energy in kinetic motions into internal heat of stars and that    otherwise, without collisions the energy in  stellar motions behaves  adiabatically%  (o worst in the case of binary heating source)
 .  However, it is generally believed that physical  collisions between stars are considered  an exotic  phenomena  that rarely happen in the universe (Binney \& Tremaine 2008), restricted to   only  be relevant  in the cores of  dense stellar configurations  like Globular Clusters systems (Portegies Zwart  et al. 1999), where it is well established that the cores of such dense stellar systems are unstable to suffer catastrophic runaway stellar collisions (Portegies Zwart \& McMillan 2002).  Numerical experiments have shown that runaway  collisions of the most massive stars could led to the formation of Intermediate-Mass Black Holes (IMBHs) in the  centers of typical globutar clusters (BH masses $\rm \sim 10^3\rm\msun$ can be build up before the first supernova explodes; Portegies Zwart \& McMillan 2002; Freitag et al. 2006a,b; Gurkan et al. 2004). Nevertheless, it is unclear what could happen in the  more extreme conditions of proto-galactic nucleus,  because of  the lack of detailed  N-body simulations that includes the effects of the higher densities and velocity  dispersions, that in addition to  gas dissipation  should  define a  density limit for NSCs before becoming globally unstable to catastrophic  stellar collisions. 

In this paper, we will study  the role of runaway stellar collisions in galactic nuclei, particularly, in the global stability of NSCs and the possible formation of MBHs. We start quantifying  the  role of collisions in NSCs in \S 2. We continue in  \S 3 proposing   an scenario for MBH formation in galactic nuclei from NSCs. Finally, we discuss the results  of the proposed scenario and its implications for the high redshift universe   in  \S 4.

\section{Quantifying %(Estimating)
 the  Role of Collisions in Nuclear Stellar Clusters %Observed Systems
  }

An order of magnitude estimate that  quantifies   the occurrence  of collisions in any   system with large number of particles, is to compute a collision timescale given by $\rm t_{coll} = \lambda/\sigma$, where $\sigma$ is the characteristic (dispersion) velocity of the system and $\lambda$ is   the particle (star) mean free path (Binney \& Tremaine 2008).   From the equation $\rm n\Sigma_o\lambda = 1$   a  mean free path can be probabilistically  defined (Landau \& Lifshitz  1980; Shu 1991), where  n is the number density of stars and $\Sigma_o$ the effective cross section,  giving      a collision rate %/frequency 
 of $\rm t_{coll}^{-1} = n \Sigma_o \sigma$. This is a  widely used definition  for collision timescale% and without assumptions
 , accurate enough for the general purpose of this paper,  for a more specific formula   that could be better to  quantify   collision rates in a particular problem see for example  Leigh et al (2014)% o Sesana)
 . Assuming that the stellar system is virialized, the dispersion velocity is $\rm \sigma = (GM/R)^{1/2}$, where M is the total mass and R the characteristic radius of the system. This result is generally valid in any stellar system in virial equilibrium and is also %approximately  
valid  for systems with a relevant dark matter component, using the empirically calibrated formula of (Cappellari  et al., 2006), where   the velocity dispersion  is $\rm \sigma = (GM/5f_{g}R)^{1/2}\approx (GM/R)^{1/2}$ for $f_g$ = 0.16 (Spergel  et al., 2003). Therefore, in any virialized stellar system the collision timescale is   given by
\begin{equation}
\rm t_{coll} = \frac{1}{n\Sigma_o} \sqrt{\frac{R}{GM}} \,\,\,\,\, .
\end{equation}

In an uniform system, composed only by solar mass stars, the number density is $\rm n = 3M/4\pi R^3M_\sun$. The  effective cross section $\rm \Sigma_o$, due to gravitational focusing, is for a solar mass star approximately  $\rm 100 \,\pi  R_\sun^2$ (i.e.    Eq 7.195 in Binney \& Tremaine 2008, with a Safronov number for solar mass stars with a $\rm \sigma \sim 100 km \,s^{-1}$; Leigh et al. 2012). Under these assumptions, neglecting radial concentrations, initial mass functions  and other %numerical
 dimensionless  factors of order unity, collisions will be relevant in the dynamics (and possibly becoming  unstable) of a given system with a  characteristic age $\rm t_H$, if its age  is comparable or longer than  the collision time, $\rm t_{coll} \leq t_H$, which is equivalent to the following condition: 
\begin{equation}
\rm  {\hat \rho}_{crit} \equiv  \left(\frac{4 \msun}{300 R_\sun^2 t_HG^{1/2}}\right)^{2/3} \leq MR^{-\frac{7}{3}}  \,\,\, ,
\label{eq2}
\end{equation}
where $\rm  {\hat \rho}_{crit}$ is a critical mass density, an intermediate density  between the  surface density ($\rm \propto R^{-2}$) and  a  volumetric one ($\rm \propto R^{-3}$).
The  largest %possible 
relevant value for  $\rm t_H$  is the age of the universe, which is of the order of  $\rm \sim 10^{10} \,years$ (Spergel  et al., 2003) that  gives a (minimal) critical mass density $\rm  {\hat \rho}_{crit} \sim 10^7  \,\msun pc^{-7/3}$, but galactic centers can be  one  order of magnitude younger ($\rm t_H \sim 10^{9} \, yr$). %The solid  blue line in Fig \ref{F1} displays the condition given by  Eq. \ref{eq2} for $\rm t_H =10^{10}$ years and the dashed blue lines, for (respectively) $\rm t_H =10^{8},  10^{6}$ and  $\rm 10^{4}$ years.   
Within geometrical factors of  order unity, such boundary is set by a combination of a fundamental  constant (G), with  typical parameters of our universe such as its current  age ($\rm t_H)$ and  the properties of the sun ($\rm \msun,\, R_\sun$), which it  is considered  to be an  average star in the Universe, defining the  critical   density of   stable stellar systems  for  our current cosmic parameters. Also, we arrive to such criterion without ad-hock assumptions, being the only assumption to be virialized, which is only a %minimum  
requirement for being   an stationary  stellar system.

%\begin{equation}
%\rm (\frac{4 m_\sun}{300 R_\sun^2 t_HG^{1/2}})^{2/3} R^{7/3} \leq M \,\,\, (OPCIONAL)
%\label{eq2}
%\end{equation}

\begin{figure}[h!]
\begin{center}
\includegraphics[width=11.9cm]{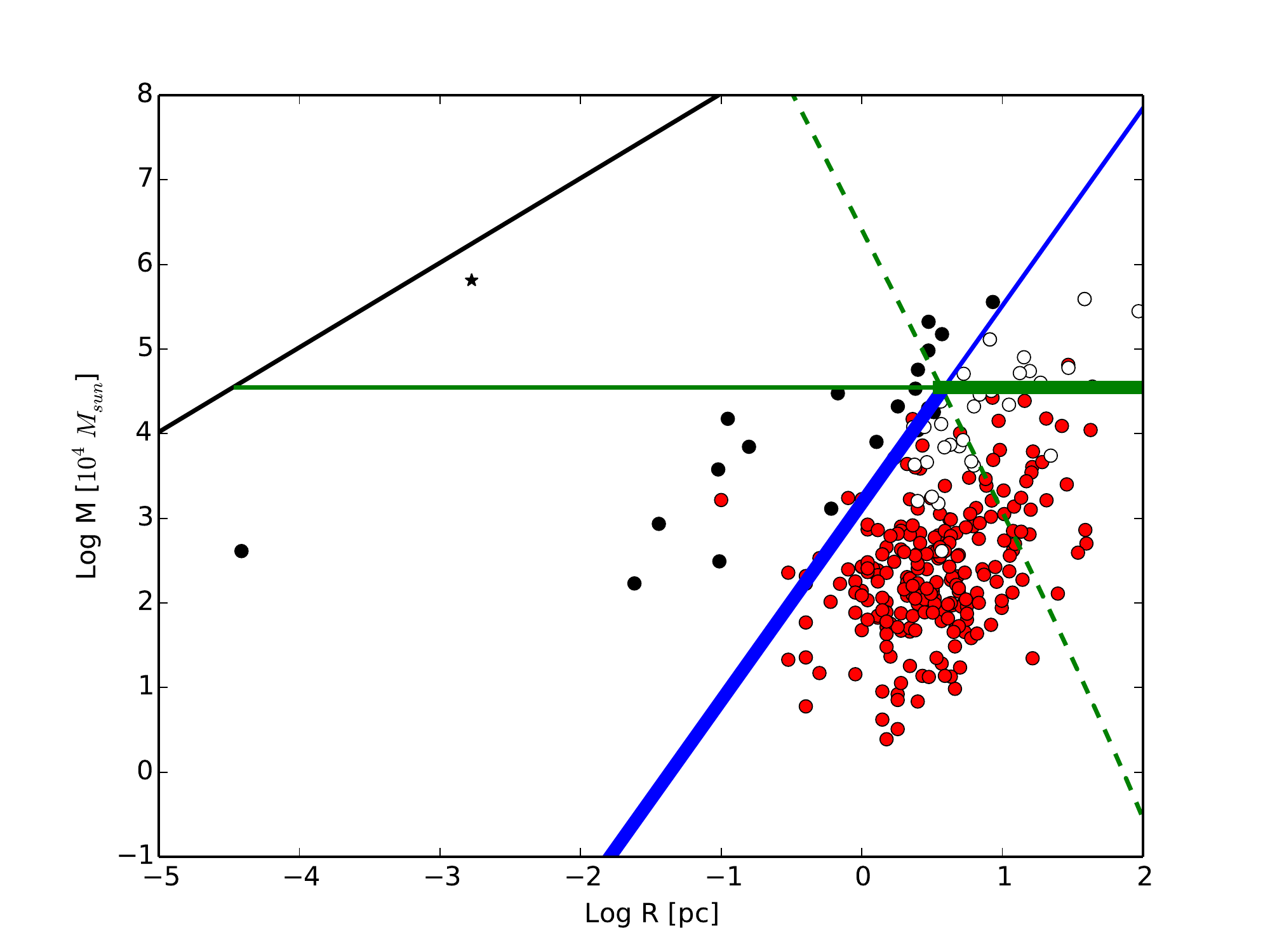}
\caption{ %SFR density as a function of the gas  surface density, divided by the 1/3 power of the radius. The symbols displayed  are: Blue triangles are galactic GMCs taken from Lada et al. (2010) and Hierderman et al. (2010). Green and red triangles are   local spiral galaxies and (U)LIRGs from Kennicutt (1998). The purple stars and cyan filled circles  are high z disks (Daddi et al. 2010a; Tacconi et al. 2010) and starburst galaxies (Genzel et al. 2010).  Pink crosses are LSB galaxies from Wyder et al. (2009). In the particular case of the galactic center we included two  estimations, both denoted by stars. 
Measured  masses and effective radius  for nuclear stellar clusters (red  circles), `well-resolved' MBHs (black circles) and `unresolved' MBHs (white circles).  The   measurement of M87's black hole shadow (The Event Horizon Telescope Collaboration 2019) is denoted by the black star, which is  the closest  to the black line that represents positions of the Schwarzschild radius as a function of  mass. The solid  blue line represents  the condition given by  Eq. \ref{eq2}  for $\rm t_H =10^{10}$ years $(\rm  
 {\hat \rho}_{crit} \sim 10^7  \,\msun \, pc^{-7/3})$% and the dashed blue lines, the same condition for respectively $\rm t_H =10^{8},  10^{6}$ and  $\rm 10^{4}$ years
 . The horizontal green line represents  the condition implied  by  Eq. \ref{eq3}  ($\rm %M_{NSC} \leq 
  \sim 3.5 \times 10^{8} \,\msun$) in order to be in agreement with  the observed scaling relation for NSCs (Leigh et al. 2012). The dashed green %purple 
line  denotes the condition given by Eq. \ref{eq4} for $\rm t_H = 10^{10} years$, which   intersects   the solid blue line  at the same critical mass %independently
 determined by the condition given by Eq. \ref{eq3}.  The positions of NSCs are      restricted within  the boundaries defined by   the collisional stable region for NSCs, denoted  by the thicker blue and green   lines.}
\label{F1}
\end{center}
\end{figure}

Figure \ref{F1}   displays the observed masses and effective radius for nuclear stellar clusters in both  late- and early-type galaxies (red  circles) taken from (Georgiev et al 2016). The solid  blue line in Fig \ref{F1} displays the condition given by  Eq. \ref{eq2}  for $\rm t_H = 10^{10}$ years%1.4 \,\,10^{10}$ years (62) 
% and the dashed blue lines, are respectively $\rm t_H =10^{8},  10^{6}$ and  $\rm 10^{4}$ years, for comparison purposes
.  The measured properties of 
NSCs (red  circles) shows a clear avoidance of the regions in which collisions could  be globally relevant in the internal dynamics of a cluster, with  collision timescales always larger than the age of the universe (right side of the solid  blue line). The only clear exception is NGC 1507, with its $\rm \geq 10^7 \,\msun$ in only 0.1 pc  of effective radius, however, this last measurement has estimated errors over 2000\% (effective radius up to 2.3 pc that moves NGC 1507 to right side of the  blue line). It is important to note that these  are average collision timescales, relevant for global stability against collisions,  that   it can be considerably shorter at the core and therefore,  these globally stable NSCs can still coexist with an unstable core which is %generally %dynamically formed 
expected  to be  triggered by Spitzer's instability (Spitzer 1969; Vishniac 1978; Portegies Zwart \& McMillan 2002).

In addition, we plot in Fig \ref{F1} the measured masses and resolution radius (=$0.5 \rm d_{resol}$, where $ \rm d_{resol}$ is the observation   spatial resolution) for MBH candidates, but we differentiate them  between `well-resolved' MBHs  with  influence radius $\rm R_{inf}$ larger than 3 spatial resolutions (black circles) and `unresolved' ones ($\rm R_{inf} < $  3 $ \rm d_{resol}$) with white circles, both from  the sample of (Gultekin  et al, 2009). %The reason is .... 
Contrarily to the case of nuclear clusters, in the case of MBH candidates we see two clear trends: the properties of  resolved MBHs are in the region that clearly passed to  the collision-dominated  regime (left side of the solid  blue line) and the unresolved ones, still avoids the collision-dominated  regime and coexist  with the NSCs. 

The trend for unresolved MBHs positions can be easily understood taking into account that the properties  of MBHs are diluted due to resolution. For the unresolved MBHs, this  means  a decrease in   densities down to values comparable to  stellar densities in the nuclear regions of their host. Therefore, the  unresolved MBHs can be taken as better  estimates of the properties of the stellar background  within $\rm R_{inf}$ than of MBHs itselfs and in some sense, they can be considered also like stellar systems. Taking this into account, the properties of NSCs clearly differs from the ones of resolved MBHs, with NSCs avoiding  the collision-dominated  region  and resolved MBHs passing  such limit, with  a sharp transition from NSCs to resolved MBHs around $\rm t_{coll}$    of the order of the age of the universe.  

 Taking also into account that for  a virialized system $\rm  R= GM/\sigma^2$, the condition given by Eq. \ref{eq2} can be rewritten %and using again virialization 
 as: 
\begin{equation}
 \rm  \sqrt{\frac{4 \, R_{\sun}}{300\,\sigma_{\sun}t_H}} \leq \frac{\msun}{M} \, \left( \frac{\sigma}{\sigma_{\sun}}\right)^{3.5} %  for a virialized system
\, , 
\label{eq3}
\end{equation}
with $\rm \sigma_{\sun}= \sqrt{\frac{G \msun }{R_{\sun}} } \sim 400 \,km \, s^{-1}$. If this condition  is combined with the empirical scaling relation that constrains the properties for observed NSCs, $\rm  \frac{M_{NSC}}{10^{6.9}\msun} = (\frac{\sigma}{128km/s})^{2.73} \sim (\frac{3\sigma}{\sigma_{\sun}})^{2.73}$ (Leigh et al. 2012), gives that NSCs will be unstable for masses larger than $\rm %M_{NSC} \geq 
\sim3.5 \times 10^{8} \msun$ (for a $\rm t_H$ again of the order of   the age of the universe). This condition is denoted by the horizontal green line in Fig. \ref{F1}, showing again good agreement with the value of the most massive NSCs. Therefore, besides these conditions being    order of magnitude estimations with %(many)
 simplifications% (in the order of geometrical factors)
 , the positions of NSCs are suggestively     restricted within  the boundaries defined by   the collisional stable region, denoted  by the thicker blue and green   lines in Fig. \ref{F1}.
 
 Although   this   trend  of  NSCs in Fig. \ref{F1} is  very suggestive, its interpretation is nevertheless   more complicated, since several additional factors needs to be taken  into account. In addition to the intrinsic idealizations of these  order of magnitude estimations, one very relevant factor is that  Fig. \ref{F1} compares the structural properties of NSCs currently observed and not at the time of formation, which can be smaller by a factor of 10 or more in radius  (Banerjee \& Kroupa 2017)% (as it should be)
, therefore,   evolutionary processes must be taken into account.  Numerical studies  of the long term evolution of NSCs (and with MBHs;  Baumgardt et al 2018; Pamanarev et al. 2019), have shown that their effective radius  indeed expands, which implies that NSCs moved from left to right in Fig. \ref{F1} as they evolve. Such expansion is expected to happen in a self-similar manner  when is only due to two-body relaxation processes ($\rm R \propto t^{2/3}$; Henon 1965;  Gieles et al. 2012;).

The two-body relaxation time  is  the relevant timescale for   %the occurrence of
 this  evolution expected in the position of NSCs in Fig. \ref{F1}, because  it quantifies     the energy exchange  by two-body scattering  and is generally expressed as  $\rm t_{relax} \approx \frac{0.1 N}{lnN} \,  t_{cross}$ (Binney \& Tremaine 2008), where  $\rm t_{cross}$ is the crossing time of the cluster and N its   total number of stars. For a virialized system, the crossing time is given by $\rm t_{cross} = R/\sigma = R^{3/2}/\sqrt{GM}$ and if is composed by solar mass stars, the total number of stars is simply N=M/$\rm \msun$. Under these  assumptions, two-body relaxation will be relevant  in the evolution of a system with a  characteristic age $\rm t_H$, if  $\rm t_H$  is comparable or longer than  the relaxation time, $\rm t_{relax} \leq t_H$, which is equivalent to the following condition: 
 \begin{equation}
\rm R  \leq \left(\frac{t_H \msun}{0.1} \, ln(M/\msun)\right)^{2/3} \,  \left(\frac{G}{M}\right)^{1/3}  \,\,\, . 
\label{eq4}
\end{equation}

The dashed green %purple 
line  in Fig. \ref{F1} denotes the condition given by Eq. \ref{eq4} for a  characteristic cluster age comparable to  age of the universe, $\rm t_H = 10^{10} years$, showing an  intersection with the solid blue line strikingly at the same critical mass %independently
 determined by the condition given by Eq. \ref{eq3}.  NSCs that  has $\rm t_{relax} \leq t_{coll}$ (with cluster mass below the intersection), could start with a collision time shorter  than the age of the universe  but its  effective radius  will  expand before collisions become important, moving  its position  the collisional stable region (right side of the solid blue line). On the other hand, if  $\rm t_{coll} \leq t_{relax}$  the NSCs will be not be able to expand before the onset of collisions, thus possibly becoming globally unstable against collisions. Therefore, this condition naturally explains the absence of NSC for masses larger than the intersection between the dashed green and  solid blue line    ($\rm M_{NSC} \gtrsim  4 \times 10^{8} \msun$). %because of being unstable against collisions
It is important to emphasize, that this  stability condition comes from   the triple equality:  $\rm t_{relax} =  t_{coll} = t_H \,(\sim 10^{10}\,yr)$.
 
Moreover,  since  the NSCs are clustered below the intersection between blue and dashed green lines in Fig. \ref{F1} ($\rm t_{relax} \leq t_{coll}$), their collision times measured today  can be assumed longer than it was at the formation of the NSCs, before their (effective) radius increased. Therefore,  the NSCs that today
displays a collision time slightly longer than the age of the universe could have had a considerably shorter collision time at formation, suggesting  that those NSCs  at  formation crossed the condition of being  in the collisional-dominated regime,   $\rm t_{coll} \leq t_{H}$, but was reversed since $\rm t_{relax} \leq t_{coll}$, meaning   that the initial conditions of galactic nuclei can indeed fulfill the conditions for instability, at least temporarily, and that many NSCs were at the edge of collisional instability  at formation.  Under  different circumstances are resolved MBHs, all in the collisional-dominated regime in Fig. \ref{F1}   ($\rm  t_{coll} \leq t_H$), independent to the relative value of  $\rm t_{relax} $, suggesting  %(at least)
 two different  channels for MBH formation at least. This fact motivates us to explore a joint MBH and NSC formation scenario in the next section, in order to explain their relative  trends  in Fig. \ref{F1}.

 %-------------------------------------------------------

%(more interesting is how close to the blue line, than to be on the right side,  meaning that most NSCs crossed the condition. They only reverse it when $t_{coll} \geq t_{relax}$, otherwise, directly collapse to a MBH. (EVIDENCIA DE  CI DENTRO DEL COLL REGIME) that the MBHs stay in the collisional-dominated regime, 
 
%Of course is idealized/simplified, more complicated possibilities/interpretations $t_{relax}$ aca porque parte del problema es el $t_{age}$ y el $\rm t_{coll}$ at formation, clarify why Hubble NATASTRON but ... To understand this suggestively result, interesting to look  $t_{relax}$ that controls the energy exchange  and mass redistribution. (i.e. after the half-mass radius increased)

%.... but should be taken as an upper limit  because  relevant  factors not taken into account, specially  the dissipation from a gaseous components that  should decrease such critical mass limit or the collision time of 2+1 scattering!!!! 

%----------------------------------------------------------

\section{An Scenario for MBH Formation  in Galactic Nuclei from  %NSCs 
Nuclear Star Clusters}

Both MBHs and  NSCs are observed to coexist in the nuclei of galaxies (Leigh et al. 2012; Georgiev et al 2016) suggesting to be a generic byproduct of their formation and evolution,  being these two completely distinct type of objects  unified   into  the terminology of being a Central Massive Object (CMO; Ferrarese  al. 2006). MBHs dominates in galaxies with masses larger than $\rm 10^{12} \msun$ and similarly  occurs  with NSCs for galaxies less massive than  $\rm 10^{10} \msun$, with both coexisting in the intermediate mass regime (Georgiev et al 2016). If  they are indeed different evolutionary stages of a common formation mechanism, the simplest interpretation of their different locations in Fig 1 is in terms of MBHs and NSCs  being CMOs with    different final fates. CMOs that are too dense to be globally  stable against stellar collisions and that also  fulfill  the condition $\rm t_{coll} \leq t_{relax}$,  thus being unable to expand before the onset of runaway collisions, it  %(may) 
will probably  collapse towards the formation of a MBH. Contrarily, this will only be the case at best in the core of less dense clusters, being globally stable in the form of a NSC, probably coexisting on its center  with a lower mass  BH formed in the  unstable core. %probably  dynamically decoupled  by Spitzer's instability).

Simulations of globular-type stellar clusters ($\rm \leq 10^{6}\,\msun$)  shows that  cores are unstable to suffer catastrophic runaway stellar collisions of massive stars due to   Spitzer's instability (Portegies Zwart \& McMillan 2002), as long as the cluster is enough massive and concentrated, with core collapse (relaxation) times less than those set by the evolution of their massive stars ($<$3-25 Myr).  A  central most massive object  is generically  formed  with efficiencies  %$\rm  M_{BH}/M_{cluster}$
  ranging from  0.1\% (Portegies Zwart \& McMillan 2002; Freitag et al. 2006a,b) up to a few percentage of the cluster mass (Sakurai et al. 2018; Reinoso et al 2018), depending on multiple physical parameters such as stellar radius, (initial) stellar mass  distribution, etc. The central most massive object is expected to have similar fate of a VMS, being typically out of thermal equilibrium, with Kelvin-Helmholtz timescale larger than  the collision timescale (Goswami el al. 2012) and also,  expected to collapse %afterwards
   to an IMBH due  to post-Newtonian instability  (Tolman 1934; Oppenheimer \& Volkoff 1939). 

 %(VMS -$>$ MBH). 

Unfortunately, direct N-body simulations  do not explore either the regime of larger clusters in the NSC  mass range ($\rm >10^{6}\,\msun$) or the   more extreme regime of globally  unstable clusters ($\rm >10^{8}\,\msun$), %$\rm t_{coll} \leq t_H$,  
not only  because  its properties are more exotic %than  those typically observed in stellar clusters 
but also, because they  are numerically much more  expensive.  The few exceptions %trace back
are  restricted  to either  Monte Carlo calculations (Sanders 1970; Gurkan et al. 2004) or self-consistent Fokker-Planck models   of galactic nuclei (Lee 1987; Quinlan \& Shapiro 1990) but their  results are already quite suggestive,  finding that in large N systems ($>10^7$ stars; which corresponds to cluster masses larger than $\rm 10^7 \msun$,  assuming  solar mass stars)  three-body binary heating is  unable to reverse core collapse before the onset of runaway collisions and then are vulnerable to a `merger  instability', which may lead to the formation of a central black hole  (Lee 1987; Quinlan \& Shapiro 1990). Since in this regime   the collision runaway started well before core collapse and  for a system  with  (initially)  equal mass stars (Lee 1987; Quinlan \& Shapiro 1990), without even requiring Spitzer's instability, it is reasonable to expect in those systems efficiencies  $\rm  M_{BH}/M_{cluster}$  higher than the  few percentage found in  N-body  simulations of globular-type stellar clusters. %Therefore,   galactic nucleus 
NSCs are indeed expected to be the most favorable places for stellar collisions  in the Universe.
 
% and  further supports the result of Fig 1, that strongly suggests that the most massive and denser  NSCs   might   exist only temporarily in the Universe%, but  they are unobservable  because of being short living entities
%, being   unstable to collapse directly to a MBH. 
  
 %Indeed, simulations of core collapse driven stellar collision in a globally stable cluster  is consistent  with efficiencies $\rm \epsilon_{BH}$ up to a few tenths? of a percentage for the most massive object (Reinoso et a. 2018; ETC). Global instability for what most of the mass ends up onto the MBH. GRAY area, GAS, etc.

More recently, Davies, Miller \& Bellovary (2011); Miller and Davies (2012) extended the  work of  Quinlan \& Shapiro (1990) using  analytical estimations  to determine the stability of NSCs against stellar collisions. In particular, Miller and Davies (2012) studied different paths  for MBH formation in NSCs and determined,  that MBHs should form in NSCs with velocity dispersions  $\gtrsim$40 km$s^{-1}$, since primordial  binaries might not provide enough  heat source via single-binary interactions that could  support core collapse as  in   Lee (1987),  Quinlan \& Shapiro (1990). They found that during or after  full core collapse, the  stars will undergo runaway collisions that produce a black hole, which will then grow via tidal disruption of stars, a process that  is expected to scale as $\rm M_{BH}(t) \sim 10^{6}\msun \, (\sigma/50km s^{-1})^{3/2}\sqrt{t/10^{10} yr}$ (Stone et al. 2017). This can be considered further evidence   to expect in galactic nuclei %those systems 
  $\rm  M_{BH}/M_{cluster}$  efficiencies higher than the  few percentage found in  N-body  simulations of globular-type stellar clusters. 
  
Nevertheless, for velocity dispersions $\gtrsim$100 km$s^{-1}$, Miller and Davies (2012) argued that NSCs will typically have too long a relaxation time for its core to collapse within  a Hubble time, a condition  equivalent to the right side of the  dashed green  displayed  in Fig. \ref{F1}. However, they missed the possible new regime  of global instability proposed here for $\rm t_{coll} \, (\leq t_H) \leq t_{relax}$, for masses larger than when the dashed green line in Fig. \ref{F1}   intersects  the solid blue line and that strikingly coincides with the maximum   mass scale with presence  of NSCs in galactic nuclei. For $\rm t_{coll} \leq t_{relax}$, relaxation processes will be unable to reverse global collapse before the onset of runaway collisions, which may lead to global collapse  onto the formation of a MBH.  
  %Davies, Miller \& Bellovary (2011) also study the role of a gaseous component in the runaway collision process, but we leave  that discussion to \S 4 in the context of the high z Universe.
%Besides the lack of detailed N-body simulations for the globally catastrophic case
Since the condition   $\rm t_{coll} \leq t_{relax}$ in Fig. \ref{F1} is computed for total NSCs quantities, is valid over the whole system and not restricted to the core, being also valid independent of different processes  favoring or quenching core collapse  analyzed by Miller and Davies (2012). 

Therefore, Fig. \ref{F1} is supporting  evidence that, in addition of NSCs being the most favorable places for stellar collisions, %therefore 
  the most massive and denser  NSCs that forms  in the Universe  might   exist only temporarily, being in principle %extremely dense stellar clusters are 
globally unstable to collapse to a MBH. This instability should be eventually triggered  by runaway stellar collisions  at some density limit, regardless if it is at the   $\rm \hat\rho_{crit}$ defined by Eq. \ref{eq2} (for $\rm t_{coll} \leq t_{relax}$) or another criterion that includes  processes  such as gas dissipation and  others not taken into account.
 It is then possible  to visualize  the following transition in the properties  of CMOs:  for objects  denser than some critical limit, %$\rm \hat\rho_{crit}$, 
which from the intersection defined by Eqs  \ref{eq2} and \ref{eq4} at $\rm t_H \sim 10^{10}\,yr$, or from Eq \ref{eq3} in order to  fulfill the observed scaling relation for NSCs (Leigh et al. 2012), seems to be  the case for $\rm M_{CMO} \gtrsim 4 \times10^8 \msun$, most of the CMO  mass will be in the form of a MBH. On the opposite mass limit, the bulk of mass in the  CMO will stay  in the stars of the NSC,  even some  cases with an undetectable MBH at its   center, with   black hole  efficiencies   $\rm \epsilon_{BH} = M_{BH}/M_{CMO}$  probably in the range of star cluster simulations from 0.1\% up to a few percent (Portegies Zwart \& McMillan 2002; Freitag et al. 2006a,b; Sakurai et al. 2018; Reinoso et al 2018) %$\sim$0.1-10\%?) 
until it approaches to a second critical mass ($\rm M_{CMO} \sim  10^7 \,\msun$; according to Lee 1987; Quinlan \& Shapiro 1990; Miller and Davies 2012; Stone et al 2017), where the black hole  efficiency should have a drastic  change, rapidly growing  towards   $\rm \epsilon_{BH}$ close to 1.

%RESUMEN de simulations Portegies Zwart \& McMillan (2002) arriba en la INTRO

%How this can proceed? 

%From N-body simulations three regimes can be seen:
 
% i) IMBH (undetected) formed at the center of NC (Portegies Zwart \& McMillan 2002) 

%ii) MBH (M$> 10^6 \msun$) at the center of NC (Scaled version of Portegies Zwart \& McMillan 2002 with growing efficiencies, less than 20\% efficiency?) 

%iii) MBH (M$> 10^{8-9?} \msun$) with no /little NC  because the efficiency at some moment dramatically increases, some process more dramatic than core collapsing. todas las estimaciones de los numeros de la Eq (Portegies Zwart \& McMillan 2002). This explains also why there is no NSC with masses $> 10^{8} \msun$.

\subsection{Testing the Scenario for  MBH Formation and  Implications for %Predicted 
Scaling Relations}

%(Modelo de eficiencias) 
The concordance of the proposed  scenario for  CMOs evolutionary paths, with the  observed relative masses in MBHs and NSCs can be easily tested. Assuming  that is the total mass  in CMOs the   mass reservoir for which competes   MBHs and NSCs in galactic nucleus, $\rm M_{CMO} = M_{NSC}+M_{BH}$, for a black hole formation efficiency  ($\rm \epsilon_{BH}$) %and a fixed  total mass  in central massive  objects
 the central black hole mass is $\rm M_{BH} = \epsilon_{BH} \,M_{CMO}$  and the mass of the surrounding 
nuclear cluster is  then $\rm M_{NSC} = (1-\epsilon_{BH}) \,M_{CMO}$,  both   related to the efficiency as $\rm \epsilon_{BH}= (1+\frac{M_{NSC}}{M_{BH}})^{-1}$, which then it  can be  directly estimated by measuring the masses $\rm M_{NSC}\, and \, M_{BH}$. Also, assuming that  is total mass  in central massive  objects $\rm M_{CMO}$   the one that correlates with the total mass of the host spheroid ($\rm M_{CMO} = \epsilon M_{sph}$, with $\epsilon\sim 0.1\%$;  Magorrian et al. 1998), it is straightforward to realize  that in both limiting cases (either only a MBH or a NSC), the observed (individual) relations are automatically fulfilled ($\rm M_{NSC} \sim \epsilon M_{sph}$ for $\rm \epsilon_{BH} \sim 0$ and $\rm M_{BH} \sim \epsilon M_{sph}$ for $\rm \epsilon_{BH} \sim 1$).
 
 \begin{figure}[h!]
\begin{center}
\includegraphics[width=11.9cm]{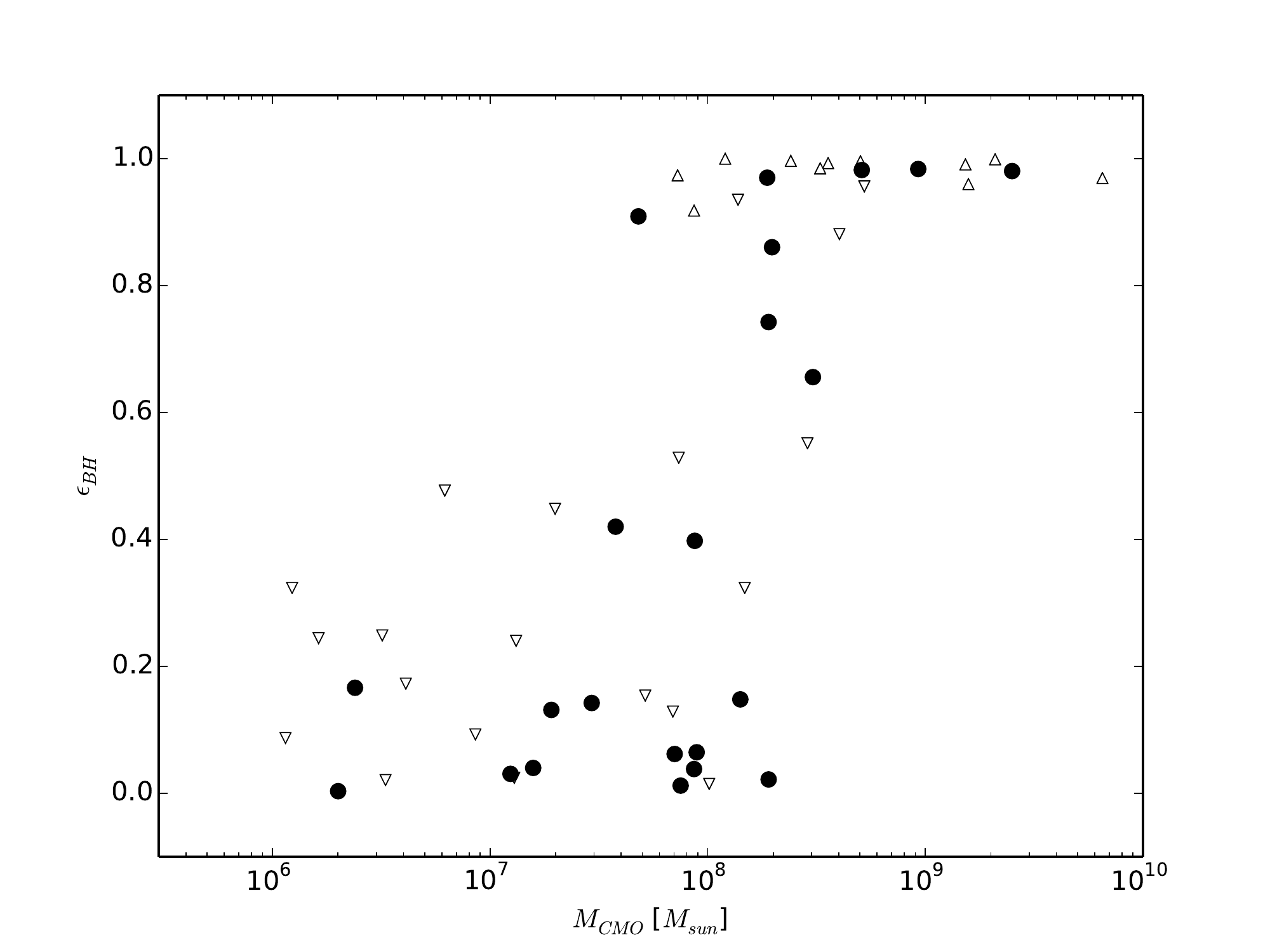}%ew.pdf}
\caption{Observed black hole formation efficiency  $\rm \epsilon_{BH}= (1+\frac{M_{NSC}}{M_{BH}})^{-1}$ as a function of  the total mass  in central massive objects  $\rm M_{CMO} = M_{NSC}+M_{BH}$, with both quantities computed using the MBHs and NSCs masses from the dataset  by Neumayer et al. (2020),   represented by   black circles%(74)
. The efficiency displayed in the figure has  two dominant values for black hole efficiencies ($\rm  \epsilon_{BH} \leq 0.15$  at $\rm M_{CMO} \leq 3\cdot10^7 \msun $ and  $\rm \epsilon_{BH} \geq 0.9$ for $\rm M_{CMO} \geq 3\cdot 10^8 \msun $) and a transition  close to a step function of the mass. Upper limits for the efficiencies  $\rm \epsilon_{BH}$ are denoted by lower triangles, while lower limits for $\rm \epsilon_{BH}$ by upper triangles, displaying the same trend of the black circles but with larger scatter at lower CMO masses%$\rm M_{CMO}$
.}
\label{F2New}
\end{center}
\end{figure}
 
Fig \ref{F2New}  displays the efficiencies  $\rm \epsilon_{BH}= (1+\frac{M_{NSC}}{M_{BH}})^{-1}$ plotted against the total mass in central massive  objects ($\rm M_{CMO} = M_{NSC}+M_{BH}$), both 
quantities computed using  measured masses from % two independent  datasets   denoted by  the white (73) and black circles (74; assuming   $\rm M_{NSC} = 10^5 \msun$ for  12 `core galaxies' with no NSC detections)
 a  dataset collected (and homogenized) recently in the    review by Neumayer et al. (2020), denoted by  the black circles on the figure. The data in Fig \ref{F2New} clearly displays three  regimes: a) $\rm \epsilon_{BH} \leq 0.15$ for $\rm M_{CMO} \leq 3 \cdot  10^7 \msun $, b) $\rm \epsilon_{BH} \geq 0.9$ for $\rm M_{CMO} \geq 3 \cdot 10^8 \msun $ and c) a transition regime between  $\rm 3 \cdot  10^7 \msun \leq M_{CMO}  \leq 3\cdot 10^8 \msun $ with rapidly growing  $\rm \epsilon_{BH}$. Upper limits for the efficiencies  $\rm \epsilon_{BH}$ are denoted by lower triangles, while lower limits for $\rm \epsilon_{BH}$ by upper triangles, displaying the same trend of the black circles, but with a larger scatter as expected for  upper and lower limits.  
 This trend is  clear and suggestively    in agreement  with the proposed formation scenario for  CMOs, with the transition  regime limited on the  expected boundaries defined by the `merger  instability' found in  Fokker-Planck models   of galactic nuclei ($\rm \gtrsim  10^7 \msun$; Lee 1987; Quinlan \& Shapiro 1990)  and the upper limit given by  the condition  $\rm t_{coll} \leq t_{relax}$  ($\rm  \gtrsim 10^8 \msun$, also in agreement with Eq \ref{eq3} to fulfill the  scaling relation for NSCs; Leigh et al. 2012). This can be contrasted, for example, with black hole efficiencies $\rm \epsilon_{BH}$ that are  randomly distributed between 0 and 1, which could be the case on a different formation scenario, where most (75\%) of the measurements should be in the interval $\rm \epsilon_{BH} = [0.15,0.9]$ and where  there is a  total absence of black circles in regimes a) and b) of Fig \ref{F2New}. %Even in the case of en independent formation, there is no reason for a void  in the interval $\rm \epsilon_{BH} = [0.2,0.9]$, which is equivalent to $\rm\frac{M_{NSC}}{M_{BH}}$ between 0.? and 10? 
Moreover, a sharp transition is also seen around  to $\rm M_{CMO} \sim 10^8 \msun$, suggesting again that this is the limit for (collision-driven) global collapse, where  most of the mass ends up into a single  MBH and that naturally  %automatically
 explains  the lack of NSCs around MBHs for  $\rm M_{BH} \gtrsim 4\cdot10^8 \msun$ (Georgiev et al 2016).

The efficiency  displayed in Fig \ref{F2New}, with two dominant  values for BH efficiencies and  a transition  with a form close to a step function of the mass, could also explain  the origin  in the change of  scaling in the M-$\rm \sigma$ relation from NSCs to MBHs, which  has been  taken as support  that MBHs and NSCs may not share a common origin (Leigh et al. 2012). The empirical evidence is that NSCs have a less steep   scaling relation $\rm M_{NSC}\propto \sigma^{2-3}$  (Leigh et al. 2012; Graham  2012), compared with  the scaling  for MBHs that  have  steeper slopes of $\rm M_{BH}\propto \sigma^{4-5}$ (Ferrarese \& Merritt 2000; Gebhardt et al. 2000). Assuming that CMOs have  a single  scaling relation originated in the galaxy formation process, for example, the scaling defined at the critical threshold   given  by Eq. \ref{eq3} ($\rm M_{CMO} \propto \sigma^{3.5}$), the step function   efficiency $\rm \epsilon_{BH}$ shown in Fig \ref{F2New} bias  the relation for  the less massive MBHs, giving a steeper slope ($>3.5$) for the  MBH scaling relation and vice versa, $\rm 1\,-\,\epsilon_{BH}$  bias  the original relation for the more massive   NSC giving  a less steep    slope ($<$ 3.5). Therefore, this naturally  reconcile  an scenario of joint formation%can have a steeper slope ($\rm M_{BH}\propto \sigma^{4-5}$)
 , with  different  M-$\rm \sigma$  scaling relation for MBHs and NSCs. %Additionally, this  automatically explains  the lack of NSCs around MBHs for  $\rm M_{BH} > 10^8 \msun$ (Georgiev et al 2016). 

 It is important to note, that alternative  scenarios  completely different  from in-situ NSC formation have been also explored  to explain NSC properties  and  the different scaling %(M-sigma)
  relations for NSCs and MBHs. For example,  a widely studied scenario is  that NSCs form via merging of   orbitally migrated globular clusters (Tremaine et al. 1976; Capuzzo-Dolcetta 1993; Leigh et al. 2015).  In those  scenarios,  tidal forces due to  the gravity  of a MBH can stop the NSC assembly by disrupting the orbitally in-falling  clusters, leading also to a maximum mass scale  for NSCs under the precence of MBHs (Antonini et al 2012, Arca Sedda et al 2015). Hybrid  models  of both  migration of stellar clusters  and in-situ formation in galactic nuclei has also been explored by means of   semi-analytical/numerical  models, which  also explains  the different  M-$\rm \sigma$ relations for NSCs and SMBHs  (%Gnedin et al 2014, 
  Arca Sedda and Capuzzo-Dolcetta 2014, Antonini et al 2015). The biggest advantage of the scenario outlined in this paper is its simplicity, giving a clear  cause for the sharp transition  in Fig. 2  for clusters  becoming globally unstable against collisions for $\rm t_{coll} \leq t_{relax}$, at the correct mass-scale for clusters with  ages comparable to the current age of the universe (Fig. 1), that in addition naturally explains  different  M-$\rm \sigma$  scaling relation for MBHs and NSCs, compared with more complex (and multifactorial) explanations in these alternative   models. 

\section{Discussion} 

%In this paper, w
We  explored  an scenario for MBH formation driven
by stellar collisions in galactic nuclei, proposing a new formation regime of global instability in NSCs triggered by runaway stellar collisions, for central massive objects that have average relaxation times longer that their collision times. For systems with ages comparable to the Hubble time, this condition is fulfilled for objects more massive than $\rm \sim 4 \times10^8\msun$, being in principle subject  to be globally unstable against stellar collisions, where  most of its mass may collapse towards the formation
of a MBH, being effectively  the fate of  failed stellar cluster. Contrarily, this will only be the case at the core of less dense central massive objects leading to the formation of MBHs with much lower efficiencies $\rm \epsilon_{BH}$%=\frac{M_{BH}}{M_{CMO}}$
. We showed that  the proposed scenario successfully explains the relative trends observed in the masses, efficiencies and scaling relations between MBHs and NSCs.

The proposed scenario      %also
 links naturally  to the fact that the existence MBHs in galaxies is intimately related with their  spheroidal/triaxial  component   (Ferrarese \& Merritt 2000; Gebhardt et al. 2000), that  is supported by random motions and  where collisions are much more frequent  compared to  the disk component of galaxies  (for a given characteristic velocity), since disks are systems that are rotationally supported and their    ordered motions  prevents collisions. %CMOs seems to be a byproduct of spheroids....
In addition,  this collision driven global instability in  extreme stellar systems  sets internally  the upper mass  limit   of NSCs around $\rm \sim 10^8 \msun$, something  needed because   at galactic scales  the study of gravitational instabilities  do not set    externally an   upper limit  for the stellar cluster masses in galactic  nuclei, since   the size of  
 the whole system  is the largest unstable wavelength (Jeans 1902). %, which  requires (to introduce) the Jeans swindle in the analysis (78). %This instability scenario also explains why there are no stellar clusters with $\rm M \geq 10^8 \msun $. 
Only when rotation becomes relevant (i.e.   in the galactic  disk), this sets a maximum mass scale for a gaseous collapsing cloud,  ranging from  $\rm M^{max}_{cloud} \sim 10^6 \msun $ for MW type disks, upto the order of  $\rm  \sim 10^8 \msun $ for ULIRGs nuclear disks  (Escala \& Larson  2008). Those massive clouds in ULIRGs are expected to migrate and runaway merge in galactic  nuclei (Elmegreen et al. 2008), again lacking of  an externally   defined upper limit.

A relevant  issue not studied in extent %detail
 in this paper is the role of gas in the formation and evolution of NSCs, particularly, in the enhancement of stellar collisions.
Although the details in the exact evolution of the   gaseous (and stellar) material funneled into galactic nuclei are still unclear under  realistic conditions, some idealized analytical estimations can  be made. For example, Davies, Miller \& Bellovary (2011) concluded that  a typical NSC at high z (mass $\rm \sim 10^6 \msun$ and size $\sim$1 pc)  can be contracted on dynamical timescales due to  gas inflow with  a mass up to  ten times heavier than the pre-existing stellar mass (according to cosmological simulations; Bellovary et al. 2011), then reach   a central density high enough for triggering  a phase of runaway collisions that could  form a MBH seed of $\rm 10^5\, \msun$ or larger.  Moreover, these high central  densities (and thus runaway collisions) can be enhanced by  the gaseous dynamical friction, that  can be an efficient process to lead the migration of additional  stellar compact remnants to galactic nuclei (Boco et al 2020).

 \begin{figure}[h!]
\begin{center}
\includegraphics[width=11.9cm]{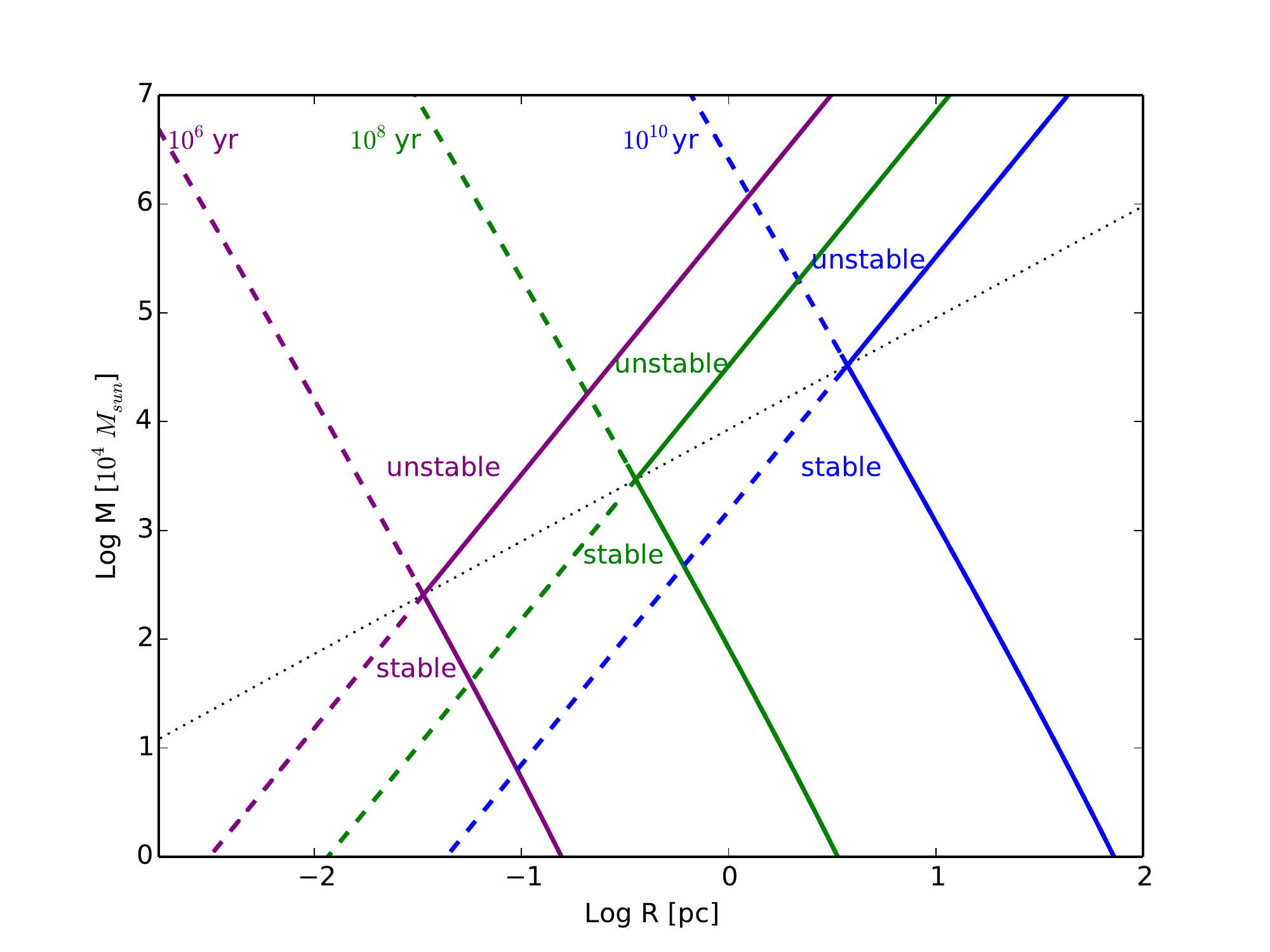}
\caption{Collision and relaxation timescales  in the mass versus   radius diagram,  for different cluster ages $\rm  t_H$:$\, 10^{10}$yr  (blue), $\rm   10^8$yr  (green) and $\rm   10^{6}$yr (purple). For each color wih a different $\rm  t_H$, clusters in the left side of the solid curves fulfill the condition $\rm t_{coll} \leq t_H$ or $\rm t_{relax} \leq t_H$, with the intersection of  solid curves dividing the stable ($\rm t_{relax} \leq t_{coll}$) and  the unstable ($\rm t_{coll} \leq t_{relax}$) regions.   The dotted black line denotes the condition $\rm t_{relax} = t_{coll}$.
}
\label{F3}
\end{center}
\end{figure}
 
 However, all these gas dynamical  processes relevant for high z (proto-)clusters act on local dynamical timescales,  short compared to the current age of the universe%($\rm \sim 10^{10}$yr)
 , therefore, is  relevant to see how the proposed  regime of collision-driven global instability can work under such shorter timescales. Fig \ref{F3} display  collision and relaxation timescales  that are equals  to cluster ages  $\rm  (t_H) \, of \, 10^{10}$yr  in blue, $\rm   10^8$yr  (green) and $\rm   10^{6}$yr (purple), which illustrates  that one order of magnitude of contraction  in effective radius, for example, due to  the gaseous  inflows expected at high z (Davies, Miller \& Bellovary 2011),  can effectively reduce in almost four orders of magnitude  the  time required  for collisions to become relevant (from $\rm t_{coll}$ being comparable to the Hubble time down to $\rm  \sim 10^{6}$yr). The dotted black line denotes the condition $\rm t_{relax} = t_{coll}$, equivalent to $\rm R/R_{\sun} = [ln(M/M_{\sun})/7.5]^{-1/2} \,\, M/M_{\sun}$, valid regardless of the cluster age. 
 
The left side of solid lines denotes  the region where 2-body relaxation or collisions can be relevant (i.e shorter than $\rm  t_H$), thus subject  to be globally  unstable to stellar collisions for $\rm t_{coll} \, (\leq t_H) \leq t_{relax}$ and globally stable for $\rm t_{relax}% \, ( \leq t_H) 
 \leq t_{coll}$.  If the discussed gas dynamical processes are able to populate clusters in the region denoted as `unstable' in green and purple on Fig \ref{F3} , they will be subject to runaway stellar collisions over the whole system on shorter timescales, possibly  leading to the formation of `naked' MBHs (i.e. without a surrounding cluster) with lower masses than currently observed in NSCs:  $\rm M_{BH}\gtrsim 3\times 10^{7}\msun \,for \, t_H \sim 10^8yr$ (green) and $\rm M_{BH}\gtrsim 3\times 10^{6}\msun \,for \, t_H \sim 10^6yr$ (purple). In the region below that threshold (intersection of solid lines), relaxation processes  will operate and the cluster will expand before the onset of runaway collisions is triggered over the whole system, being such collisions restricted to the unstable core and thus a central MBH  is possibly formed that   coexists with a surrounding NSC.  
 
% But if the protocluster is able to form, new plot predicting high z? 
 
%  UNSTABLE (Globally): $\rm t_{coll} (\leq t_H) \leq t_{relax}  $
 % STABLE (Globally): $\rm t_{relax} ( \leq t_H) \leq t_{coll} $

 %Therefore, %m
 Most probably 
 the  extremely dense, purely gas-free  NSC %idealized hypothesized
 discussed    in this paper rarely exists in the early Universe and most often% (especially   in the early Universe)
 , an unstable    NSC  will collapse  as a whole during its formation, %onto a MBH
  before  evaporating its gaseous  envelope, as might be suggested by the multiple stellar populations seen in the surviving NSCs (Nishiyama \& Schodel 2012). 
  %  . Also, we have   neglected  relevant secular processes such as  the role of preexisting  MBHs binaries  in the dynamics  and formation of galactic nuclei (80-82).
On the other hand, if the formation of this  surviving NSCs %dense 1e8 Msol clusters 
was occurring at high redshift, these would likely be observable with James Webb Space Telescope (JWST). Since the progenitors of  globular clusters are already expected to be detected  at high-z (3 $\lesssim$ z $\lesssim$ 8; Renzini 2017), these most massive progenitors of NSCs  should be more likely to be   detected, as the number of clusters that could be detected scales linearly with their mass (Renzini 2017). %If your suggestion includes a "prompt" formation channel, you should consider this as a potential observable. 

Certainly, more  realistic simulations (with and without gaseous components) are  needed to set the open issues, but the  absence  of NSCs in the collision-dominated regime, with the sharp transition seen at the boundaries of the unstable  region of Fig 1, suggests  that  the fate of the unstable ones  is  unavoidably  collapsing  onto a  MBH. Therefore, besides all these uncertainties, the results in this work can be taken as supporting  evidence that the collapse leading to MBH formation is most probably triggered by runaway  collisions, than by suppressing  fragmentation on smaller scales or  alternatively, by the runaway growth of a preferred IMBH on cosmological timescales. Also, this collision-driven BH formation is a process that could  happen even in the the earliest epochs of the Universe (Korol et al. 2019), without imposing   strict constraints on cosmological timescales. % like the growth of an  IMBH seed.    

%3.- PARRAFO VMS/FINAL COLLAPSE/eLISA.  Besides the  supporting  observational evidence in favor of stellar collision-driven  MBH formation, The final collapse onto a MBH will have direct implications for LISA.
   
Because  it is hard to constrain enough  MBH formation  thru  direct  observations  of such objects by  traditional electromagnetic detections, in addition of  having more complex  and realistic simulations,  definite answers will probably come from direct observations of the final collapse by gravitational-wave observatories  such as LISA (Amaro-Seoane  et al. 2013).  In the complex collision-driven collapse scenario described in this letter, it is hard that the final collapsing  VMS will be close to  spherically or axially symmetric, therefore, it is expected to be at least bar-shaped and  most probably, even more irregular and a gravitational wave signal it is expected from galactic centers at the moment of MBH formation (Rees 1984),  that will be detectable in the LISA band out to high redshift (Sun et al. 2017).

\end{document}